\documentclass[a4paper, 11pt]{article}

\usepackage[dvips]{graphicx}
\usepackage{epsf}
\usepackage{latexsym}
\usepackage{amssymb}
\usepackage{a4wide}

\title{Approximating the zero-coupon bond price \\
in a general one-factor model with constant coefficients}

\author{Be\'ata Stehl\'ikov\'a\footnote{Department of Applied Mathematics and Statistics, Faculty of Mathematics, Physics and Informatics, Comenius University, 842 48 Bratislava, Slovakia. \newline
E-mail: \texttt{stehlikova@pc2.iam.fmph.uniba.sk}}}

\date{}

\begin{document}

\maketitle

 \begin{abstract}
We consider a general one-factor short rate model, in which the instantaneous interest rate is driven by a univariate diffusion with time independent drift and volatility. We construct recursive formula for the coefficients of the Taylor  expansion of the bond price and its logarithm around $\tau=0$, where $\tau$ is time to maturity. We provide numerical examples of convergence of the partial sums of the series and compare them with the known exact values in the case of Cox-Ingersoll-Ross and Dothan model.
 \end{abstract}

\section{Introduction}

Interest rate is a rate charged for the use of the money. 
 As an~example we show the evolution of several interest rates on the market in Figure \ref{fig:intro} above. Figure \ref{fig:intro} below displays the interest rates with different maturities (so called term structures) at a given day.

\begin{figure}[ht] 
  \begin{center}
 \includegraphics[width=0.9 \textwidth]{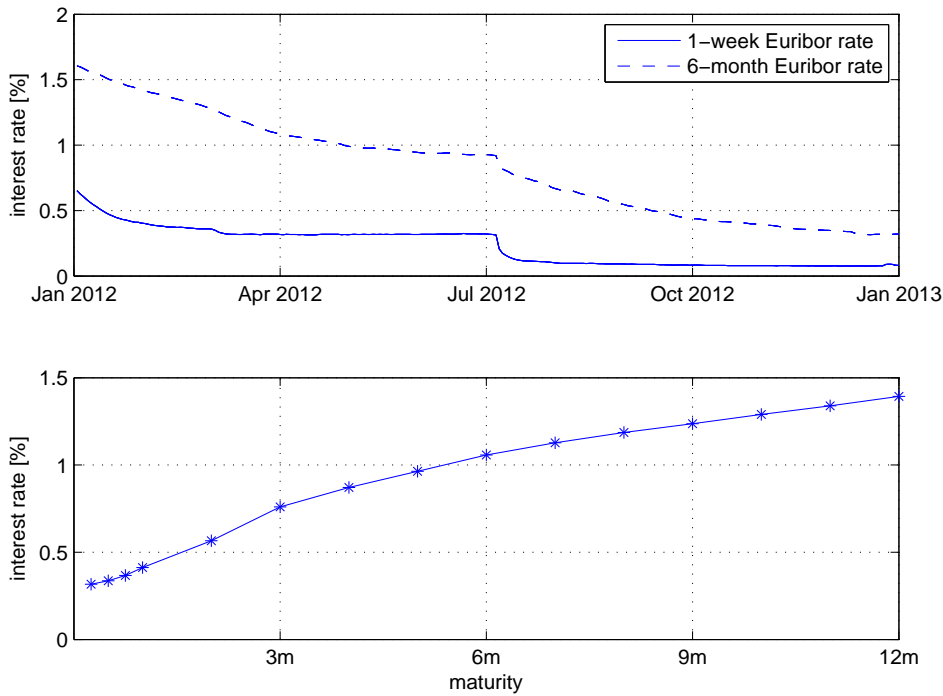} 
 \end{center}  
  \caption{{\footnotesize Example of the time evolution of interest rates (above) and term structure of interest rates (below).}}
 \label{fig:intro}
 \end{figure}

Short rate models are based on specifying the evolution of the instantaneous interest rates (also called short rate) in terms of a stochastic differential equation. The remaining interest rates are then determined by the bond prices. After specifying the so called market price of risk, the price of a bond is a solution to the partial differential equation (PDE). Alternatively, the short rate evolution can be given in so called risk neutral measure. In particular, if the short rate evolves in the risk neutral measure according to the stochastic differential equation
$$dt = \mu(r,t) dt + \sigma(r,t) dw,$$
where $w$ is a Wiener process, then the price $P=P(t,r)$ of the zero-coupon bond at time $t$ when the current level of the short rate is $r$, is a solution to the PDE 
\begin{equation}
\frac{\partial P}{\partial t} + \mu(r,t) \frac{\partial P}{\partial r} + \frac{1}{2} \sigma^2(r,t) \frac{\partial^2 P}{\partial r^2} - r P=0.
\label{eq:pde1}
\end{equation}
The PDE (\ref{eq:pde1}) holds for all $r>0$ and $t \in (0,T)$, where $T$ is the maturity of the bond. Morover, since the bond pays its principal value at maturity, which can be normalized to be 1, the function $P$ moreover satisfies the terminal condition $P(r,T)=1$ for all $r>0$. The interest rates $R(t,r)$ is then given by the formula
\begin{equation}
P(t,r)=e^{-R(t,r)(T-t)},\;\textrm{ i.e., } R(t,r)=-\frac{\log P(t,r)}{T-t}. \label{eq:vztah-p-r}
\end{equation}
 See, e.g., \cite{brigo-mercurio} or \cite{kwok} for more details on interest rates modelling.

Solution to the bond pricing PDE (\ref{eq:pde1}) is known only i special cases, such as Vasicek model \cite{vasicek} or Cox-Ingersoll-Ross (CIR) model \cite{cir}. Sometimes, there is a closed form solution known but it is complicated, for example the bond price in Dothan model \cite{dothan} involves special functions and two-dimensional integration. In general, however, there no closed form solution avalible. Therefore, the bond prices are computed using Monte Carlo simulations, numerically solving the PDE (\ref{eq:pde1}) or 
by looking for analytical approximative formulae. We follow the last approach in this paper.

In papers \cite{choi-wirjanto}, \cite{stehlikova-sevcovic}, \cite{stehlikova}, the 
Chan-Karolyi-Longstaff-Sanders (CKLS) model \cite{ckls}
$$dr = (\alpha + \beta r) dt + \sigma r^{\gamma} dw$$
and analytical approximate formulae for bond prices were considered. The time independence of the coefficients of the model enables us to write the bond price in terms of the time to maturity $\tau=T-t$. The common feature of these approximations is  that their order of accuracy can be expressed in the form
\begin{equation}
\ln P^{ap}(\tau,r) - \ln P (\tau,r) = c(r) \tau^{\omega} + o(\tau^{\omega}) \label{eq-order}
\end{equation}
as $\tau \rightarrow 0^+$, where $P$ is the exact bond price and $P^{ap}$ is the proposed approximation. The relation (\ref{eq-order}) asserts that the Taylor series of $\ln P^{ap}$ and $\ln P$ coincide up to the certain order. In particular, in \cite{stehlikova-sevcovic} it has been shown that for the formula from  \cite{choi-wirjanto} the relation (\ref{eq-order}) holds with $\omega=5$  and an improvement leading to $\omega=7$ has been derived. In \cite{stehlikova} a simple formula with $\omega=4$ has been proposed. These results suggest that the Taylor expansion (either of the price itself and its logarithm) could be a good approximation too. A numerical illustration is presented in Table \ref{tab:motivation-bond} and Table \ref{tab:motivation-rates}.

{\footnotesize
\begin{table}[ht]
\begin{center}
\begin{tabular}{cccccc}
\hline 
maturity & exact & Choi-Wirjanto & Taylor, order 4 & Taylor, order 5 & Taylor, order 6 \\ 
\hline \hline 
0.25 & 0.987567 & 0.987567 & 0.987567 & 0.987567 & 0.987567 \\ 
0.5 & 0.975273 & 0.975273 & 0.975273 & 0.975273 & 0.975273 \\ 
0.75 & 0.963120 & 0.963121 & 0.96312 & 0.963120 & 0.96312 \\ 
1 & 0.951115 & 0.951115 & 0.951115 & 0.951115 & 0.951115 \\ 
1.5 & 0.927559 & 0.927559 & 0.927559 & 0.927559 & 0.927559 \\ 
2 & 0.904626 & 0.904629 & 0.904627 & 0.904626 & 0.904626 \\ 
2.5 & 0.882334 & 0.882342 & 0.882336 & 0.882333 & 0.882334 \\ 
3 & 0.860691 & 0.860709 & 0.860696 & 0.860688 & 0.960691 \\ 
4 & 0.819367 & 0.819435 & 0.819382 & 0.819348 & 0.819368 \\ 
5 & 0.780631 & 0.780813 & 0.780662 & 0.780565 & 0.780638 \\ 
\hline 
\end{tabular} 
\caption{{\footnotesize Zero coupon bond prices in the CIR model with parameters  $\alpha=0.00315$, $\beta=-0.0555$ and $\sigma=0.0894$ (taken from \cite{choi-wirjanto}), $r=5 \%$: exact solution, approximation from \cite{choi-wirjanto}  and   Taylor expansions of the exact solution around $\tau=0$. }}
\label{tab:motivation-bond}
\end{center}
\end{table}}

{\footnotesize
\begin{table}[ht]
\begin{center}
\begin{tabular}{cccccc}
\hline 
maturity & exact & Choi-Wirjanto & Taylor, order 4 & Taylor, order 5 & Taylor, order 6 \\ 
\hline \hline 0.25 & 5.00425 & 5.00425 & 5.00425 & 5.00425 & 5.00425 \\ 
0.5 & 5.00766 & 5.00766 & 5.00766 & 5.00766 & 5.00766 \\ 
0.75 & 5.01024 & 5.01023 & 5.01023 & 5.01024 & 5.01024 \\ 
1 & 5.01202 & 5.01201 & 5.01201 & 5.01202 & 5.01202 \\ 
1.5 & 5.01328 & 5.01324 & 5.01327 & 5.01329 & 5.01328 \\ 
2 & 5.01167 & 5.01152 & 5.01163 & 5.01169 & 5.01167 \\ 
2.5 & 5.00739 & 5.00704 & 5.00729 & 5.00745 & 5.00739 \\ 
3 & 5.00065 & 4.99994 & 5.00046 & 5.00078 & 5.00064 \\ 
4 & 4.98059 & 4.97852 & 4.98014 & 4.98115 & 4.98054 \\ 
5 & 4.95306 & 4.94839 & 4.95227 & 4.95474 & 4.95288 \\ 
\hline 
\end{tabular} 
\caption{{\footnotesize Interest rates in the CIR model with parameters  $\alpha=0.00315$, $\beta=-0.0555$ and $\sigma=0.0894$ (taken from \cite{choi-wirjanto}), $r=5 \%$: exact solution, interest rates computed from the bond price approximation from \cite{choi-wirjanto}  and  from  Taylor expansions of the exact solution around $\tau=0$. }}
\label{tab:motivation-rates}
\end{center}
\end{table}}


The paper is organized as follows. In section 2 we
show how to construct the Taylor expansion of the bond price and its logarithm for
a general one-factor model with constant coefficients
\begin{equation} 
dr = \mu(r) dt + \sigma(r) dw \label{eq:sde2} 
\end{equation}
In  section 3 we test them in the case of CIR model (where the explicit solution is known) and Dothan model (where the approximate solution with a given accuracy has been. In the final section we give some concluding remarks.

\section{The Taylor expansions}

In this sections we derive the serie expansions of the bond price and its logarithm. Considering the logarithm of the price is useful when studying the relative error in the prices or the absolute error in the interest rates, see the formula (\ref{eq:vztah-p-r}).

\subsection{Computing the Taylor expansion of the  bond price}

Recall that the bond price $P=P(t,r)$ satisfies the PDE (\ref{eq:pde1}). Since he coefficients of the diffusion (\ref{eq:sde2})
do not depend on time $t$, using the transformation $\tau=T-t$ we can write the bond price as $P=P(\tau,r)$ and it satisfies
\begin{equation}
-\frac{\partial P}{\partial \tau} + \mu(r) \frac{\partial P}{\partial r} + \frac{1}{2} \sigma^2(r) \frac{\partial^2 P}{\partial r^2} - r P=0  \label{eq:pde-tau}
\end{equation}
for all $r>0$, $\tau \in (0,T)$ and the initial condition $P(0,r)=1$ for all $r>0$.

We construct the Taylor expansion of $P(\tau,r)$ around $\tau=0$ in the form 
\begin{equation}
P(\tau,r) = \sum_{j=0}^{\infty} c_j(r) \tau^j. \label{eq-P-2}
\end{equation}
From the initial condition $P(0,r)=1$ we see that $c_0(r)=1$ for all $r>0$. Inserting the series expansion (\ref{eq-P-2}) into (\ref{eq:pde-tau})
we get
$$-\left[ \sum_{k=0}^{\infty} k c_k(r) \tau^{k-1} \right] + \mu(r)\left[ \sum_{k=0}^{\infty} c'_k(r) \tau^{k} \right]
 + \frac{1}{2} \sigma^2(r)  \left[ \sum_{k=0}^{\infty}  c''_k(r) \tau^{k} \right] - r \left[ \sum_{k=0}^{\infty} c_k(r) \tau^{k} \right] =0.$$
Comparing the coefficients at $\tau^j$ for $j=0,1,2,\dots$ we get
$$-(k+1)c_{k+1}(r) + \mu(r)c'_k(r) + \frac{1}{2} \sigma^2(r) c''_k(r) - r c_k(r) =0.$$
Now we can recursively compute the coefficients $c_{k+1}$:
$$c_{k+1}(r) = \frac{1}{k+1} \left[
\mu(r)c'_k(r) + \frac{1}{2} \sigma^2(r)  c''_k(r)  - r c_k(r) \right].
$$
Computing $c_1, c_2,\dots,c_J$ in this way, we truncate the serie and obtain an approximation
$$
  P(\tau,r) \sim \sum_{j=0}^{J} c_j(r) \tau^j.
$$

\subsection{Computing the Taylor expansion of the  logarithm of the bond price}

We proceed in the same way as in the case of the bond price in the previous subsection.
Let us denote the logarithm of the bond price as $f$, i.e.,
$$f(\tau,r)= \log P(\tau,r).$$
After a simple transformation of the PDE (\ref{eq:pde-tau}) we obatin the PDE satisfied by $F$ which reads as 
\begin{equation}
-\frac{\partial f}{\partial \tau} + \frac{\sigma(r)^2}{2}  \left[ \left( \frac{\partial f}{\partial r} \right)^2 +  \frac{\partial^2 f}{\partial r^2} \right] + \mu(r) \frac{\partial f}{\partial r} - r =0.
\label{eq-f}
\end{equation}

In the same way as in the case of the price itself, we consider the logarithm of the bond price. We construct the expansion around $\tau=0$ in the form 
\begin{equation}
f(\tau,r) = \sum_{j=0}^{\infty} c_j(r) \tau^j. \label{eq-f2}
\end{equation}
From the initial condition $f(0,r)=0$ we see that $c_0(r)=0$ for all $r>0$. Inserting the series expansion (\ref{eq-f2}) into (\ref{eq-f})
we get
$$-\left[ \sum_{k=0}^{\infty} k c_k(r) \tau^{k-1} \right] + \mu(r)\left[ \sum_{k=0}^{\infty} c'_k(r) \tau^{k} \right]
 + \frac{1}{2} \sigma^2(r) \left[ \sum_{k=0}^{\infty} c'_k(r) \tau^{k} \right]^2 $$ $$ + \frac{1}{2} \sigma^2(r) \left[ \sum_{k=0}^{\infty} c''_k(r) \tau^{k} \right] - r \tau^0 =0.$$
Comparing the coefficients at $\tau^0$ we get
$$-1 c_1(r) + \mu(r)c_0'(r) + \frac{1}{2} \sigma^2(r) c'_0(r)^2 + \frac{1}{2} \sigma^2(r) c''_0(r) - r =0$$
and by substituting $c_0(r)=0$ for all $r>0$ we obtain
$c_1(r)=-r.$
Comparing the coefficients at $\tau^j$ for $j=1,2,\dots$ we have
$$-(k+1)c_{k+1}(r) + \mu(r)c'_k(r) + \frac{1}{2} \sigma^2(r) \sum_{i=0}^k c'_i(r) c'_{k-i}(r) + \frac{1}{2} \sigma^2(r) c''_k(r)=0.$$
Now we can recursively compute the coefficients $c_{k+1}$:
$$c_{k+1}(r) = \frac{1}{k+1} \left[
\mu(r)c'_k(r) + \frac{1}{2} \sigma^2(r) \sum_{i=0}^k c'_i(r) c'_{k-i}(r) + \frac{1}{2} \sigma^2(r) c''_k(r) \right].
$$
Computing $c_2,\dots,c_J$ in this way, we truncate the serie and obtain an approximation
$$
\ln P(\tau,r) \sim \sum_{j=0}^{J} c_j(r) \tau^j.
$$

\section{Numerical results}

We test the Taylor expansion approximation in the models, where either the exact solution is known, or a numerical solution with a given precision is available. We also check the convergence, as more Taylor terms are added.

\subsection{Cox-Ingersoll-Ross model}

Firstly, we consider the CIR model \cite{cir}, in which the short rate evolves in the risk neutral measure according to the stochastic differential equation 
$$dr = (\alpha + \beta r)dt + \sigma \sqrt{r} dw.$$
The bond price   can be expressed in the closed form by a simple formula, cf. \cite{cir}.

Setting $\mu(r)=\alpha+\beta r$ and $\sigma(r)=\sigma \sqrt{r}$ into the general formulae derived in the previous section results in the coefficients of the bond price and its logarithm respectively.
The first coefficients for the bond price\footnote{all the coefficients used in the computations are available from the author upon request; we omit them here for the space reasons since they are much longer for the higher orders} are given by
\begin{eqnarray}
c_0(\tau,r)&=&0, \nonumber \\
c_1(\tau,r)&=&-r,\nonumber \\
c_2(\tau,r)&=&\frac{1}{2}(\alpha + \beta r - r^2), \nonumber \\
c_3(\tau,r)&=&\frac{1}{6} \left(  -\beta(\alpha + \beta r) + r \sigma^2 \right), \nonumber \\
c_4(\tau,r)&=&\frac{1}{24} \left( 3 \beta r \sigma^2 + (\alpha+\beta r) (\sigma^2 - \beta^2) \right), \nonumber \\
c_5(\tau,r)&=&\frac{1}{120} \left(
r \sigma^2 (7 \beta^2 - 4 \sigma^2) + \beta (\alpha  + \beta r)(4 \sigma^2 - \beta^2)
\right). \nonumber
\end{eqnarray}

\begin{figure}[!ht]
 \centerline{
 \includegraphics[width=0.8\textwidth]{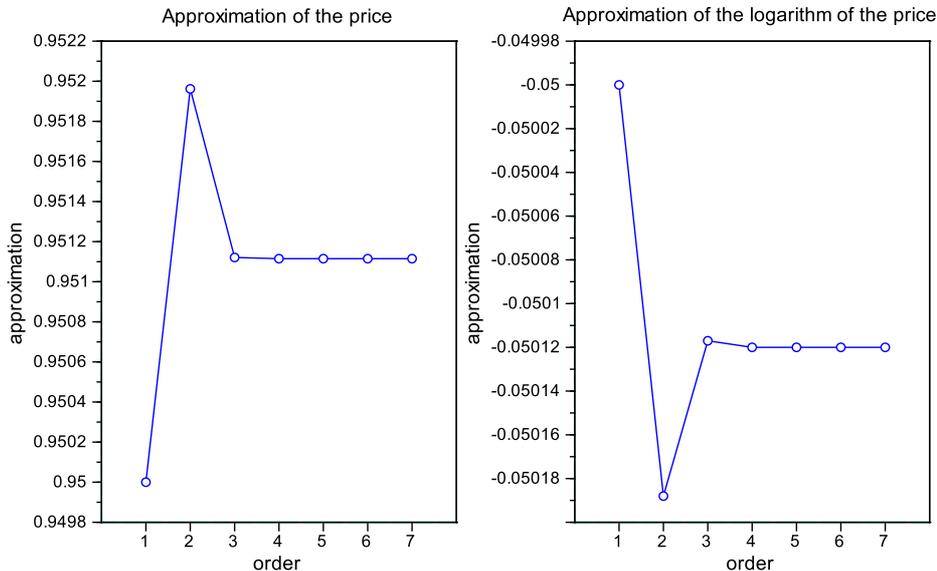} 
  }
\caption{{\footnotesize  CIR model with parameters  $\alpha=0.00315$, $\beta=-0.0555$ and $\sigma=0.0894$ - bond price (left), logarithm of bond price (right) with maturity $\tau=1$ - convergence of the Taylor approximations.}}
\label{fig-cir-1}
\end{figure}

We assume the same set of parameters as in the introduction, i.e., $\alpha=0.00315$, $\beta=-0.0555$ and $\sigma=0.0894$. Table \ref{tab-cir-1}  and Figure \ref{fig-cir-1}  show the partial sums for $T=1$ and $r=0.05$, together with the exact value of the bond price and its logarithm. Note that since  we take $T=1$,  the yield equals to the log price taken with the positive sign and we can also observe the convergence of the yields.

\begin{table}[!ht]
\centering
\begin{tabular}{ccc}
order $J$ &  approximation    \\
\hline \hline
0 & 1.000000 \\ 
1 & 0.950000  \\  
2 & 0.951062  \\ 
3 & 0.951121  \\ 
4 & 0.951115  \\ 
5 & 0.951115  \\ 
6 & 0.951115  \\ 
7 & 0.951115  \\ 
\hline
\end{tabular} 
\hskip 0.2cm
\begin{tabular}{cc}
order $J$ & approximation \\ 
\hline \hline
0 &  \phantom{-}0.000000 \\ 
1 & -0.050000 \\  
2 & -0.050188  \\ 
3 &  -0.050117 \\ 
4 &  -0.050120 \\ 
5 & -0.050120  \\ 
6 &  -0.050120 \\ 
7 &  -0.050120\\ 
\hline
\end{tabular} 

\caption{{\footnotesize  CIR model with parameters  $\alpha=0.00315$, $\beta=-0.0555$ and $\sigma=0.0894$ - bond price (left), logarithm of bond price (right) with maturity $tau=1$ - convergence of the Taylor approximations.}}
\label{tab-cir-1} 
\end{table}

\subsection{Dothan model}

The Dothan model \cite{dothan} assumes that the short rate in the risk neutral measure follows the stochastic differential equation
$$dr = \mu r dt + \sigma r dw.$$
The zero-coupon bond in the Dothan model has the explicit solution, but it is computationally complicated (cf. \cite{brigo-mercurio}).
Therefore, we use the Dothan bond prices computed in \cite{hansen-jorgensen} for which the error estimate is available. They are accurate to the given four decimal places.

Setting $\mu(r)=\mu r$ and $\sigma(r)=\sigma r$ into the general formulae derived in the previous section results in the coefficients for the price and its logarithm\footnote{they are again available from the author upon the request}.
In the numerical experiments we use the values from \cite{hansen-jorgensen}. The authors price zero coupon bonds which pays 100 USD at maturity $T$ (hence its price is 100 times the value considered so far), if  the short rate evelves according to the Dothan model with $\mu=-0.005$, $\mu=0.005$ and $\sigma^2=0.01$, $\sigma^2=0.02$, $\sigma^2=0.03$.  The initial value of the short rate is $r=0.035$. Using their iterative algorithm, for $\tau=1,2,3,4,5,10$ they obtain the accuracy to four decimal places for all combinations of parameters and in several cases  also  for higher maturities.

\begin{figure}[!ht]
 \centerline{
 \includegraphics[width=0.8\textwidth]{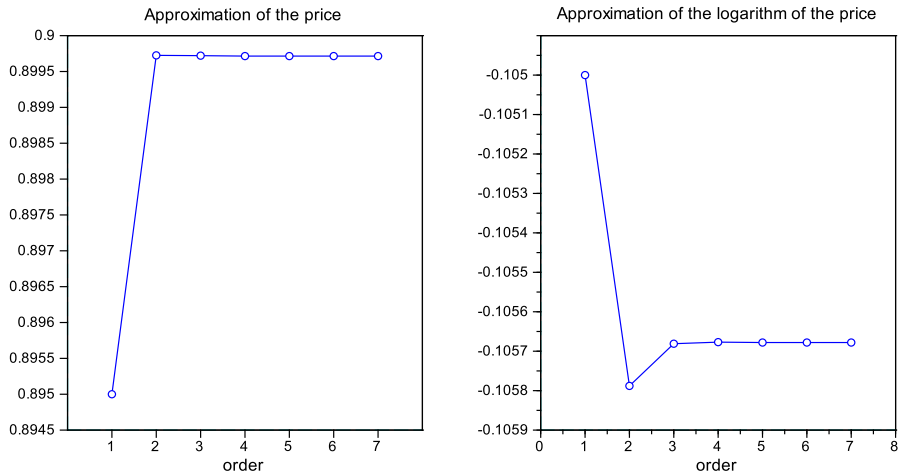} 
  }
\caption{{\footnotesize  Dothan model with parameters  $\mu=0.005$, $\sigma^2=0.02$ - bond price (left), logarithm of bond price (right) with maturity $\tau=3$ - convergence of the Taylor approximations.}}
\label{fig-dothan-1}
\end{figure}

Firstly we show an example of comvergence of partial sums for the parameter values $\mu=0.005$ and $\sigma=0.02$ and maturity $\tau=3$ in Table \ref{tab-dothan-1} and Figure \ref{fig-dothan-1}. 
Afterwards, we use both the values from \cite{hansen-jorgensen} to test our approximation for a wider range of parameters and maturities, as shown in Table  \ref{tab-dothan-2}.

\begin{table}[!ht]
\centering
\begin{tabular}{cc}
$J$ & approximation \\ 
\hline \hline
0 & 1.000000 \\ 
1 & 0.895000  \\  
2 & 0.899725  \\ 
3 & 0.899721  \\ 
4 & 0.899715 \\ 
5 & 0.899715  \\ 
6 & 0.899715  \\ 
7 & 0.899715  \\   
\hline
\end{tabular} 
\hskip 0.2cm
\begin{tabular}{cc}
$J$ & approximation \\ 
\hline \hline
0 &  \phantom{-}0.000000 \\ 
1 & -0.105000 \\  
2 & -0.105788 \\ 
3 &  -0.105681 \\ 
4 &  -0.105677\\ 
5 & -0.105678  \\ 
6 &  -0.105678 \\ 
7 &  -0.105678 \\ 
\hline
\end{tabular} 
\caption{{\footnotesize  Dothan model with parameters  $\mu=0.005$, $\sigma^2=0.02$ - bond price (left), logarithm of bond price (right) with maturity $\tau=3$ - convergence of Taylor approximations. }}
\label{tab-dothan-1} 
\end{table}

\begin{table}[t]
\centering
\begin{tabular}{ccccccc}
parameters & $\tau$ &  Taylor, J=3 &  Taylor, J=5 &  Taylor, J=7 & exact \cite{hansen-jorgensen}  \\ 
\hline \hline
$\mu=0.005$, $\sigma^2=0.01$& 1 & 96.5523 &96.5523   & 96.5523 & 96.5523 \\ 
 & 2 & 93.2082 & 93.2082   & 93.2082 & 93.2082 \\ 
& 3 & 89.9666 & 89.9663   & 89.9663 & 89.9663 \\ 
& 4 & 86.8260 & 86.8251   & 86.8251 & 86.8251 \\ 
& 5  & 83.7852 & 83.7830   & 83.7830 & 83.7830 \\ 
&  10 & 70.0312  & 69.9977   & 69.9982 & 69.9982 \\ 
\hline 
$\mu=0.005$, $\sigma^2=0.02$ & 1 & 96.5525 & 96.5525   & 96.5525 & 96.5525 \\ 
 & 2 &  93.2099 & 93.2098   & 93.2098 &  93.2098 \\  
& 3  & 89.9721 & 89.9715   & 89.9715 &  89.9715 \\  
& 4  & 86.8391 & 86.8370   & 86.8370 & 86.8370  \\
& 5  & 83.8362 & 83.8056   & 83.8057 & 83.8057\\
&  10  & 70.4396 & 70.1530   & 70.1551 &  70.1551 \\
\hline
$\mu=0.005$, $\sigma^2=0.03$ & 1 & 96.5527 & 96.5527   & 96.5527 & 96.5527 \\ 
 & 2 &  93.2115 & 93.2113   & 93.2113 &  93.2113 \\  
& 3  & 89.9776 & 89.9767   & 89.9767 &  89.9767 \\  
& 4  & 86.8521 & 86.8491  & 86.8491 & 86.8491  \\
& 5  & 83.8362 & 83.8287   & 83.8287 & 83.8287 \\
&  10  & 70.4396 & 70.3112   & 70.3151 &  70.3151 \\
\hline
\end{tabular} 
\caption{{\footnotesize  Bond prices in the Dothan model with indicated parameters and maturities, and the initial value of the short rate 
 $r_0=0.035$ - comparison of Taylor approximation with exact values }}
\label{tab-dothan-2} 
\end{table}

\section{Conclusions}
We considered the Taylor expansion for bond prices and their logarithms in a general one factor model with time independent parameters. The expansion is considered with respect to time to maturity $\tau$ around $\tau=0$. The coefficients of the Taylor expansions can be computed recursively and expressed in a closed form. Since the prices are expanded into series around $\tau=0$, the most precise resuts are expected to be obtained ofr small $\tau$. However,  numerical examples dealing with Cox-Ingersoll-Ross and Dothan models show that we are able to obtain very precise approximations also for higher values of $\tau$.

\section*{Acknowledgments}
This research was supported by VEGA 1/0747/12 grant.

\end{document}